\documentclass[aps,preprintnumbers,amsmath,amssymb,prd,superscriptaddress,longbibliography,twocolumn]{revtex4-2}
\usepackage{graphicx}
\usepackage{dcolumn}
\usepackage{color}
\usepackage{caption}
\usepackage{subcaption}
\usepackage{feynmp}
\usepackage{feynmp-auto}
\usepackage{float}
\usepackage{systeme,mathtools}
\usepackage{stmaryrd}
\usepackage{easyReview}
\usepackage{tikzsymbols}

\usepackage{ragged2e}
\DeclareCaptionJustification{fulljustified}{\justifying}
\def\comment#1{}

\newcommand{\nc}{\newcommand}
\nc{\beq}{\begin{eqnarray}}
	\nc{\eeq}{\end{eqnarray}}
\nc{\scs}{\scriptstyle}
\nc{\setval}{\fmfset{wiggly_len}{3mm} \fmfset{arrow_len}{1.5mm}
	\fmfset{arrow_ang}{13} \fmfset{dash_len}{1.5mm}\fmfpen{0.125mm}
	\fmfset{dot_size}{2thick}}

\usepackage{bm,latexsym,mathrsfs,enumerate,color}
\usepackage[mathcal]{euscript}
\usepackage[breaklinks=true,unicode=true,urlcolor = blue,colorlinks = true,citecolor = blue,linkcolor = blue]{hyperref}
\usepackage{todonotes}
\usepackage{wrapfig}

\renewcommand{\vec}[1]{\bm{#1}}

\def\slashchar#1{\setbox0=\hbox{$#1$}           
	\dimen0=\wd0                                 
	\setbox1=\hbox{/} \dimen1=\wd1               
	\ifdim\dimen0>\dimen1                        
	\rlap{\hbox to \dimen0{\hfil/\hfil}}      
	#1                                        
	\else                                        
	\rlap{\hbox to \dimen1{\hfil$#1$\hfil}}   
	/                                         
	\fi}                                         %

\DeclareMathAlphabet\mathbfcal{OMS}{cmsy}{b}{n}

\begin{document}
	
\title{Competing Magnetic Orders and Continuously Varying Exponents in $U(1)$ Spin Liquids}

\author{Vira Shyta}
\affiliation{Max Planck Institute for the Physics of Complex Systems, N\" othnitzer Strasse 38, 01187 Dresden, Germany}

\author{Flavio S. Nogueira}
\affiliation{Institute for Theoretical Solid State Physics, IFW Dresden, Helmholtzstr. 20, 01069 Dresden, Germany}

\date{Received \today}

\begin{abstract}
	Unequal couplings to a common emergent gauge field provide a continuous way to tune competing magnetic instabilities in an effective description of $U(1)$ spin-liquid mixtures. We study this mechanism in quantum electrodynamics in 2+1 dimensions with two families of four-component Dirac fermions. Combining a leading large-$N$ treatment with Schwinger–Dyson equations, we obtain chiral condensates representing Néel order in the two sectors. At fixed fermion number, the coupling ratio $\kappa=e_2^2/e_1^2$ controls which condensates develop. Depending on $N$, the phases with a single condensate are connected through either coexistence or a finite interval in which both fermion families remain gapless. The mass-generation thresholds are reciprocal, and the fermion gaps close with Miransky essential singularities. The renormalization group flow has a circular line of fixed points parametrized by the invariant ratio $\kappa$. At the calculated order, gauge-invariant staggered magnetization correlations have continuously varying exponents in the gapless regime, while the exponent for correlations between the two sectors is independent of $\kappa$. These results connect the competition between magnetic orders to a continuous family of critical theories and identify magnetic correlation signatures of the intervening gapless regime.
\end{abstract}

\maketitle

{\it Introduction}--- Quantum electrodynamics (QED) is one of the most precise theories developed so far. Its asymptotic behavior and renormalization properties are well understood \cite{zuber2012quantum}. Its renormalization group (RG) $\beta$ function features a Gaussian fixed point \cite{zuber2012quantum,zinn2021quantum} and thus the theory is infrared (IR) free, in contrast to non-Abelian theories, which are ultraviolet (UV) free 
 \cite{Gross-Wiclzek_PhysRevLett.30.1343,Politzer_PhysRevLett.30.1346}. Thus, despite its extraordinary predictive power, QED fails to be an asymptotically free theory. This fact led in the past to serious doubts about the consistency of the theory due to the presence of so-called Landau ghosts (or poles) \cite{zuber2012quantum}. However, the asymptotic behavior of QED is sharply different in spacetime dimensions lower than $d=4$. For instance, in 2+1 dimensions QED is UV free and IR interacting. The lower dimensionality also changes the context for the experimental relevance. In fact, QED in 2+1 dimensions had several different incarnations within the condensed matter physics context \cite{Marston_PhysRevB.39.11538,Herbut_PhysRevB.66.094504,Franz_PhysRevB.66.054535,Hermele_PhysRevB.70.214437,Nogueira-Kleinert_PhysRevLett.95.176406,Hermele_PhysRevB.77.224413,Sondre_PhysRevB.111.075158}. Perhaps the most famous example of a condensed matter system featuring Dirac fermions in 2+1 dimensions is graphene \cite{Castro-Neto_RevModPhys.84.1067,Gusynin-Sharapov-Carbotte,GorbarSharapov2013}. Graphene electrodynamics is particularly interesting, since it features four-component Dirac spinors and $4\times 4$ Dirac gamma matrices, just like QED in 3+1 dimensions. This  permits to define a $\gamma_5$-like matrix and therefore a chiral symmetry in 2+1 dimensions \cite{Herbut_PhysRevLett.97.146401}. Unlike the case where the number of dimensions of spacetime is even, the $\gamma_5$-like matrix is not given in terms of a product of the other gamma matrices. Microscopically, this chiral symmetry is a sublattice symmetry. Gapping graphene would break the chiral symmetry, similarly to standard, four-dimensional massless QED.

The type of chiral symmetry arising in graphene electrodynamics has been known for a long time in massless (2+1)-dimensional QED, or QED3 \cite{Pisarski_PhysRevD.29.2423,Appelquist_PhysRevD.33.3704,Appelquist_PhysRevLett.60.2575,Nash_PhysRevLett.62.3024}. A QED3 defined in terms of two-component spinors does not possess the chiral symmetry and thus allows for a massive term.  The latter, however, breaks time-reversal symmetry ($T$) and parity ($P$). In Euclidean spacetime, this amounts to just $P$, since time plays the role of an additional spatial coordinate. This version of QED3 features the so-called parity anomaly, which is the statement that one cannot keep intact the original classical discrete symmetries \textit{and} gauge invariance of the massless theory at the quantum level. In other words, for $P$ to be enforced, one spoils the gauge invariance when quantum fluctuations of the Fermi fields are accounted for \cite{Redlich_PhysRevLett.52.18}. In contrast, in QED3 with four-component spinors, the parity anomaly cancels and the mass term would break chiral symmetry instead. In this case, we have two $4\times 4$ matrices, $\gamma_3$ and $\gamma_5$, that anticommute with the Dirac $\gamma$ matrices $\gamma_0$, $\gamma_1$, and $\gamma_2$. Provided the fermions are massless, the chiral symmetries, $\psi\to e^{i\theta_{3}\gamma_{3}}\psi$ and $\psi\to e^{i\theta_{5}\gamma_{5}}\psi$, hold \cite{Pisarski_PhysRevD.29.2423,Appelquist_PhysRevD.33.3704}. Assuming there are $N$ spinors available, it has been known for a long time that chiral symmetry in QED3 is spontaneously broken if $N$ does not exceed a certain critical value $N_c$ that typically lies within the interval $(3,4)$ \cite{Appelquist_PhysRevLett.60.2575,Nash_PhysRevLett.62.3024,Gusynin_PhysRevD.94.125009}.

The mass gap resulting from chiral symmetry breaking (CSB) exhibits a behavior reminiscent of the Berezinskii-Kosterlitz-Thouless (BKT) transition \cite{Berezinskii1971,Kosterlitz1973}, where $N$ plays a role similar to the temperature in the two-dimensional $XY$ model. In such a system, the critical temperature represents an essential singularity in the gap, and the same happens with $N_c$ in QED3. In the QED3 context, this BKT-like behavior is known as Miransky scaling \cite{MIRANSKY1980421} and the physical mechanism behind it is quite different. The main concept in play here  was initially formalized by introducing the notion of conformal phase transition \cite{Gusynin_PhysRevD.58.085023}, dealing with situations where the conformal symmetry is broken at a non-perturbative level. This was further developed into closely related concepts of walking \cite{Walking-JHEP} and conformality loss \cite{Kaplan_PhysRevD.80.125005}. The basic idea can be understood with the following paradigmatic toy example often employed in the literature. Suppose we have an RG flow described by the equation, $dg/d\ln\mu=\alpha-\alpha_*-(g-g_*)^2$, where $\alpha$ is some invariant parameter and $\alpha_*$ and $g_*$ are given real numbers. The  fixed points are given by $g_\pm=g_*\pm\sqrt{\alpha-\alpha_*}$, where  $g_+$ is UV stable and  $g_-$ is IR stable. When $\alpha\geq \alpha_*$, the fixed points are real. For $\alpha<\alpha_*$, they become complex and the conformality is lost. The solution of the RG equation in this regime reveals the Miransky scaling, $\mu_{\rm IR}/\mu_{\rm UV}\sim\exp(-{\rm const}/\sqrt{\alpha_*-\alpha})$, for the ratio between the IR and UV scales. This slow RG flow is characteristic of walking.  
In QED3, the parameter tuning the transition, similarly to $\alpha$ in our toy example, is the number of fermion components $N$.  But unlike $\alpha_*$, the critical value $N_c$ does not follow from the solution of the RG equation, being rather obtained from an approximated solution of the Schwinger-Dyson (SD) equation \cite{Appelquist_PhysRevD.33.3704}.

Depending on the condensed matter system at hand, the chiral condensate in QED3 can take on different physical meanings. In graphene, it represents an exciton condensate, and the resulting phase structure is one of a semimetal-insulator transition as the strength of the Coulomb interaction is varied \cite{KHVESHCHENKO2004323,Gorbar_PhysRevB.66.045108,Son_PhysRevB.75.235423}. In $U(1)$ spin liquids, it describes the onset of antiferromagnetic order \cite{KIM1999130,Rantner-Wen_PhysRevB.66.144501}. However, in the latter case this occurs by varying the number of components $N$ corresponding to an $SU(N)$ Heisenberg antiferromagnet where the spin liquid phase appears in the large $N$ limit \cite{Marston_PhysRevB.39.11538}, mirroring exactly the earlier analysis of the chiral condensate in QED3 \cite{Appelquist_PhysRevLett.60.2575}. 

In this Letter, we investigate competing N\'eel orders in $U(1)$ spin liquids by allowing two inequivalent families of Dirac fermions to couple with different strengths to the same emergent photon. Each family can develop a chiral condensate, while both contribute to screening of the common gauge field. Their relative coupling therefore provides a continuous parameter for changing the balance between the two ordering tendencies at fixed total fermion number. We formulate this setting in QED3 with gauge couplings $e_1$ and $e_2$, and show that the ratio $\kappa=e_2^2/e_1^2$ controls two reciprocal thresholds for dynamical mass generation. Within a large-$N$ SD treatment, the phases with a single condensate are connected either through a coexistence region or through a finite interval where both fermion sectors remain gapless, depending on $N$. The fermion gaps close with Miransky scaling at the corresponding boundaries. The RG analysis of the massless theory identifies $\kappa$ as an invariant labeling a circle of IR fixed points. We further determine the gauge-independent anomalous dimensions of the mass operators and the associated staggered-magnetization correlations, revealing continuously varying exponents in the gapless regime. These results connect the imbalance between the gauge couplings among fermion sectors to both the competition between magnetic orders and the scaling of physical correlations. 

{\it Competing chiral orders}--- We now turn to the main focus of our investigation. To begin with, we consider two families of massless Dirac fields $\psi_i=[\psi_{i,1} \dots \psi_{i,N/2}]^T$, $i=1, 2$, with $N$ even. We assume that all components of $\psi_i$ are four-component spinors, so that a chiral symmetry can be defined in 2+1 dimensions, similarly to graphene. We consider a QED theory with the following Euclidean Lagrangian, 
\begin{equation}
	\label{Eq:L}
	\mathcal{L}=\sum_{i=1,2}\bar{\psi}_{i}(\slashchar{\partial}-ie_i\slashchar{a})\psi_{i}+\frac{1}{2}(\epsilon_{\mu\nu\lambda}\partial_\nu a_\lambda)^2,
\end{equation}
that features different gauge couplings for the fields  $\psi_{1}$ and $\psi_{2}$.  This can be regarded as a spin liquid mixture where the Fermi velocities are different, causing the effective gauge couplings, $e_1^2$ and $e_2^2$, to differ for distinct Dirac particles.

Assuming a large $N$ limit with $Ne_i^2$ kept fixed, the leading order vacuum polarization is given by, 
\begin{equation}
	\label{Eq:VP}
	\Pi(p)=\frac{N(e_1^2+e_2^2)}{16|p|}.
\end{equation}
We see that the photon self-energy $\Sigma_{\mu\nu}(p)=\Sigma_a(p)(\delta_{\mu\nu}-p_\mu p_\nu/p^2)$, $\Sigma_a(p)=p^2\Pi(p)$, behaves in two spatial dimensions in the same way as the scalar potential self-energy in graphene. Indeed, $\Sigma_a(0,\vec{p})=N(e_1^2+e_2^2)|\vec{p}|/16$, while in graphene the self-energy associated with scalar potential $\phi$ is given by $\Sigma_\phi(0,\vec{p})=ng^2|\vec{p}|/8$ \cite{Gorbar_PhysRevB.66.045108, Son_PhysRevB.75.235423}. Hence, the static Coulomb potential in graphene and QED3 has a long-distance behavior $\sim 1/r$.

As the photon self-energy dominates over the Maxwell term in the IR, the SD equation in the leading order takes the form, 
\begin{equation}
	\label{Eq:SD}
	\Sigma_i(p)=\frac{32e_i^2}{N(e_1^2+e_2^2)}\int\frac{d^3k}{(2\pi)^3}\frac{\Sigma_i(k)}{[k^2+\Sigma_i^2(k)]|k+p|}.
\end{equation}
Using the standard procedure to convert the SD equation to a differential equation and subsequently solving it approximately \cite{Appelquist_PhysRevD.33.3704,Nash_PhysRevLett.62.3024}, yields the spontaneous gap generation, 
\begin{equation}
	\label{Eq:Gap}
	\Delta_i=\Sigma_i(0)=\Lambda \exp\left[-\frac{2\pi}{\sqrt{\frac{64e_i^2}{\pi^2N(e_1^2+e_2^2)}-1}}\right].
\end{equation}

\begin{figure}
	\begin{minipage}{0.49\textwidth}
		\includegraphics[width=0.95\textwidth]{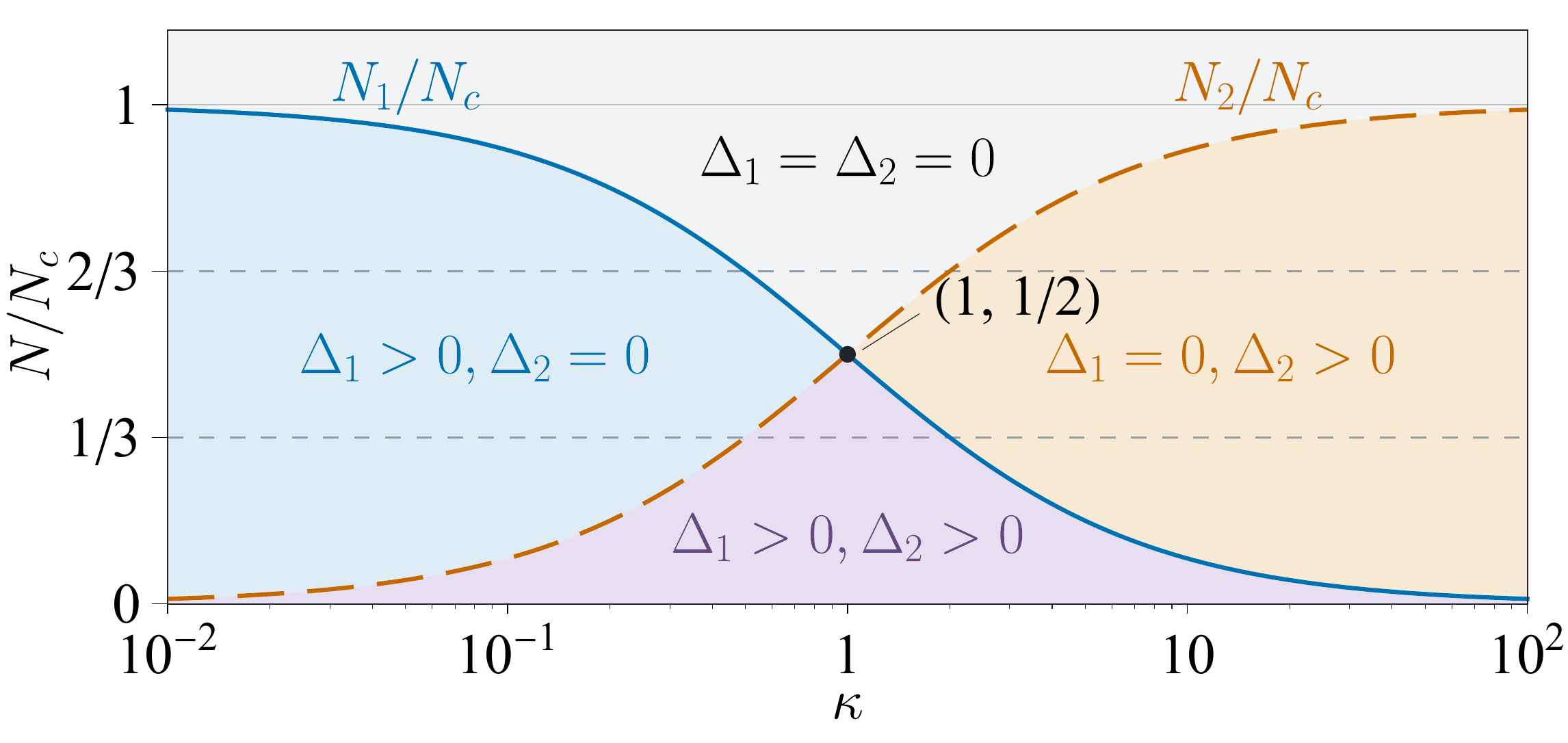}
		\captionsetup[sub]{width=0.85\linewidth}
		\subcaption{}
	\end{minipage}%
		\vspace{0.05cm}
	\begin{minipage}{0.24\textwidth}
	\includegraphics[width=0.95\linewidth]{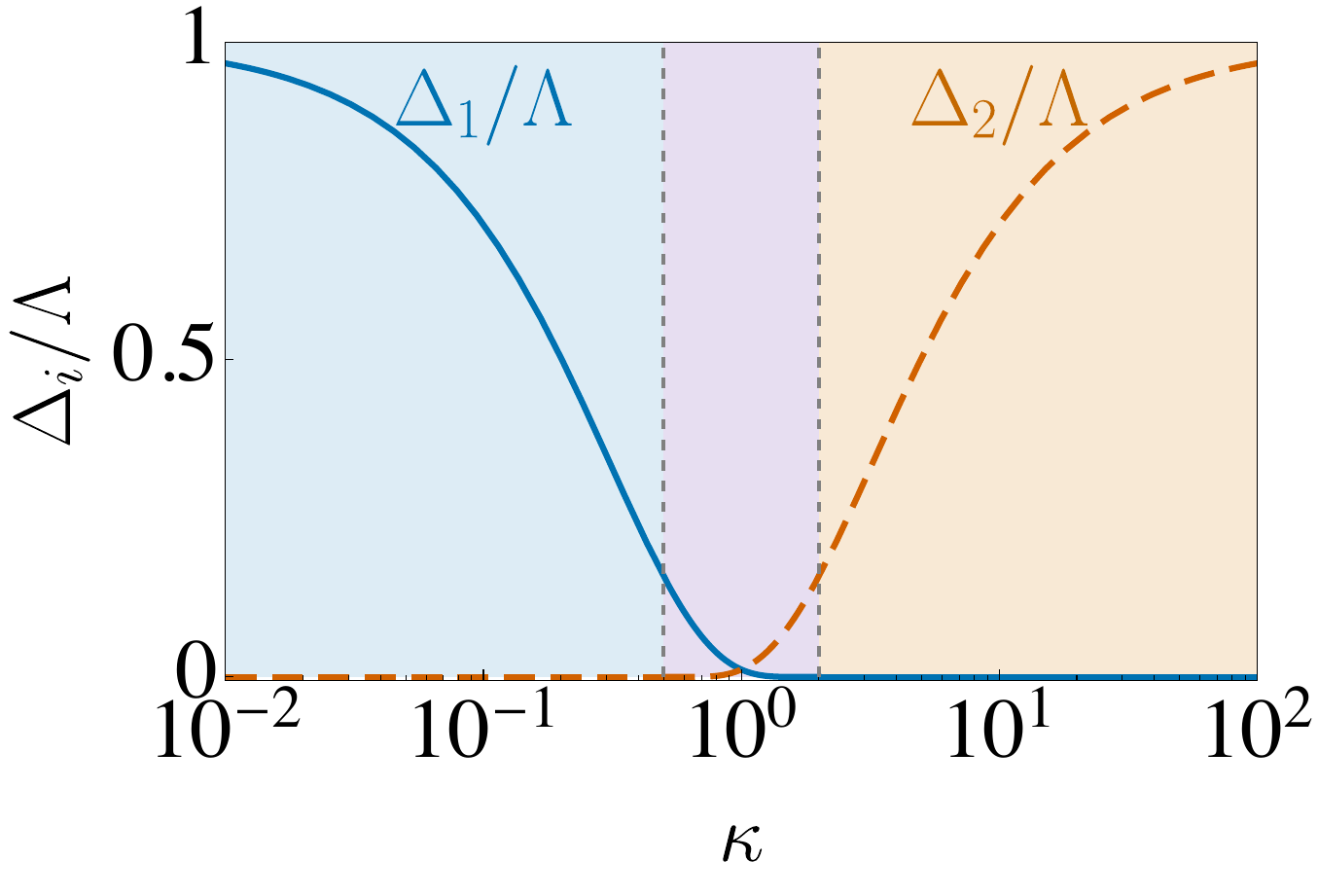}
			\captionsetup[sub]{width=0.75\linewidth}
	\subcaption{}
	\label{Fig:fig1panelb}
	\end{minipage}%
	\begin{minipage}{0.24\textwidth}
		\includegraphics[width=0.95\linewidth]{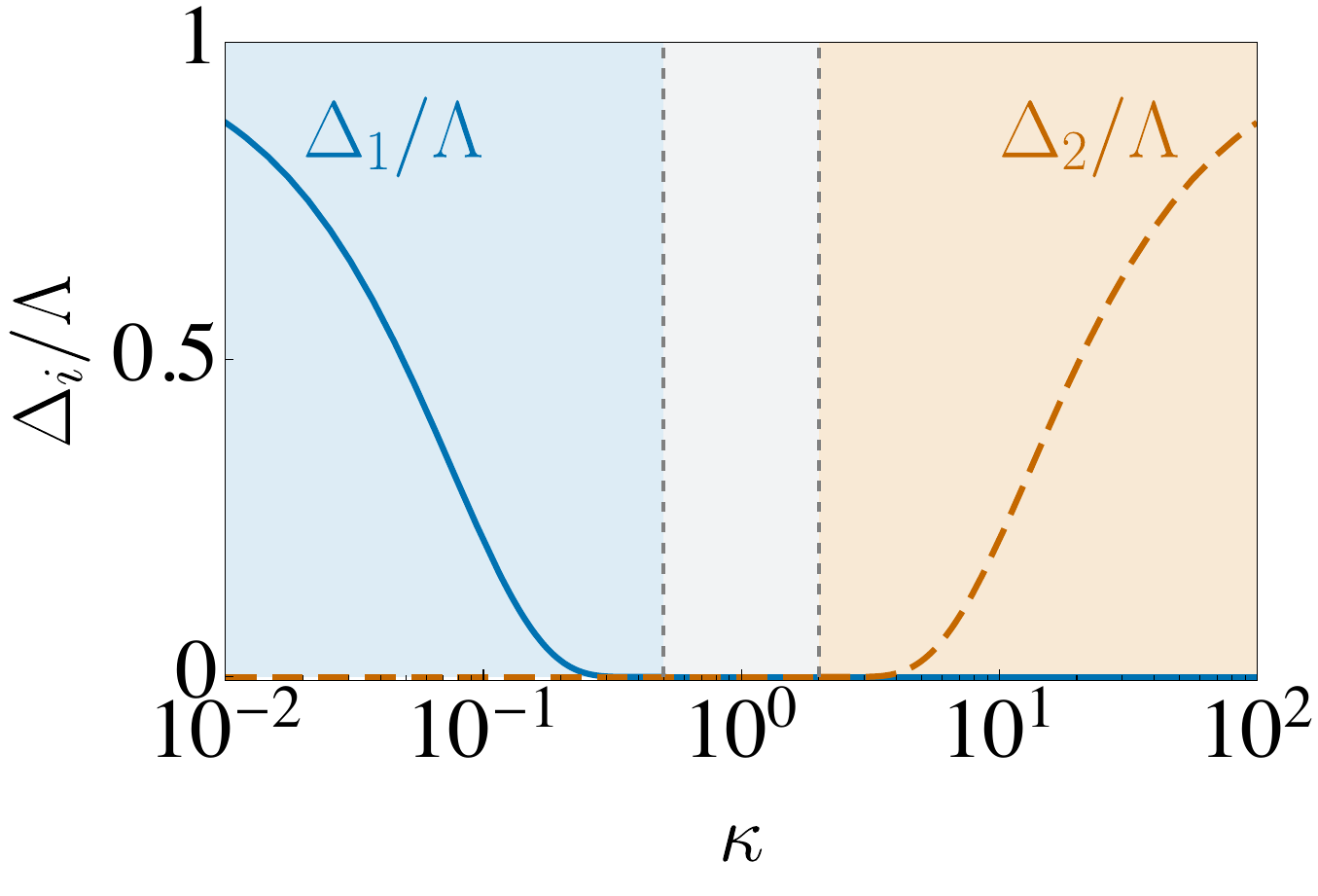}
			\captionsetup[sub]{width=0.75\linewidth}
	\subcaption{}
	\label{Fig:fig1panelc}
	\end{minipage}
	\caption{Phase diagram illustrating the behavior of the gaps as a function of the gauge couplings' ratio  $\kappa$. In all three panels the regions are color-coded as follows: (i) $\Delta_1>0$ and $\Delta_2=0$ (blue); (ii) $\Delta_2>0$ and $\Delta_1=0$ (orange); (iii) both gaps are nonzero (lavender); (iv) both gaps vanish (light gray). Panel (a) shows a phase diagram with all possible gap configurations dependent on  $\kappa$ and the number of fermion components $N$. The solid blue curve and the dashed orange curve correspond to  $N_1(\kappa)$ and $N_2(\kappa)$ of Eq.~\eqref{Eq:crit-curves-n}, respectively, normalized by $N_c=64/\pi^2$.  These critical curves intersect at $\kappa=1$, $N=N_c/2$. 		Panels (b) and (c) depict the gaps given by Eq.~\eqref{Eq:Gap} as  functions of $\kappa$  for $N/N_c=1/3$  and $N/N_c=2/3$, respectively [see the dashed light gray lines in panel (a)]. The gaps are normalized by their maximum values, e.g. $\Delta_{1\text{max}}=\Delta_1(\kappa\to 0)$.
		In panel~(b), there is an intermediate coexistence phase, where both gaps, $\Delta_1$ (solid line) and $\Delta_2$ (dashed line), are nonvanishing. This corresponds to the case where $\kappa_{c1}>\kappa_{c2}$, with the critical values, $\kappa_{c1}$ and $\kappa_{c2}$, defined in Eq.~\eqref{Eq:kappa1}. Panel (c), on the other hand, features a gapless phase since  $\kappa_{c2}>\kappa_{c1}$.}
\label{Fig:PhaseDiagram}
\end{figure}

The two gaps are not always coexisting. The fermion $\psi_1$ can be gapped, while $\psi_2$ is not, and vice versa. To see this, we introduce the tuning parameter $\kappa=e_2^2/e_1^2$ and rewrite the gaps above as $\Delta_i=\Lambda\exp(-2\pi/\alpha_i)$, where, 
\begin{equation}
	\label{Eq:alpha1}
	\alpha_1=\sqrt{\frac{64}{\pi^2N(1+\kappa)}-1}, 
\end{equation}
\begin{equation}
	\label{Eq:alpha2}
	\alpha_2=\sqrt{\frac{64\kappa}{\pi^2N(1+\kappa)}-1}. 
\end{equation}
The gap behavior is defined not only by the number $N$ of fermions but also by the ratio $\kappa$. The ensuing phase diagram is depicted in Fig. \ref{Fig:PhaseDiagram}. Let us explore it in more detail.
 
From Eqs.~\eqref{Eq:alpha1} and \eqref{Eq:alpha2}, we obtain the critical values of $\kappa$, 
\begin{equation}
	\label{Eq:kappa1}
	\kappa_{c1}=\frac{64}{\pi^2N}-1, \qquad \kappa_{c2}=1/\kappa_{c1},
\end{equation}
such that $\Delta_1\neq0$ whenever $\kappa< \kappa_{c1}$ and $\Delta_2\neq0$ for $\kappa> \kappa_{c2}$.
To ensure $\kappa_{c1}$ and $\kappa_{c2}$ are both positive, $N$  has to be smaller than the critical value $N_c=64/\pi^2\approx 6.48$.
Setting $\alpha_i^2=0$ gives the critical curves, 
\begin{equation}
	\label{Eq:crit-curves-n}
	N_1(\kappa)=\frac{N_c}{1+\kappa},\qquad
	N_2(\kappa)=\frac{N_c\kappa}{1+\kappa},
\end{equation}
below which the respective gaps are finite. The interplay between $N$ and $\kappa$ in the phase diagram is captured by panel (a) of  Fig.~\ref{Fig:PhaseDiagram}. The most interesting aspect of the phase diagram is that a gap coexistence is conditional.
It  is possible when $N<N_c/2\approx3.24$ and requires $\kappa_{c2}<\kappa<\kappa_{c1}$ (Fig.~\ref{Fig:fig1panelb}). On the other hand, if $N$ is such that $N_c/2<N<N_c$, the phase is gapless for $\kappa$ in the closed interval $[\kappa_{c1},\kappa_{c2}]$ (Fig.~\ref{Fig:fig1panelc}). At $N=N_c/2$, the interval collapses to $\kappa=1$, i.e. $e_1=e_2$; this is consistent with the usual QED3, where $N$ is the only parameter driving the phase transition, with the gap closing for $N>N_c/2$ (as we see from Fig.~\ref{Fig:fig1panelc}, this still occurs to a limited extent  when $\kappa\in[\kappa_{c1},\kappa_{c2}]$).  Finally,   for $N\ge N_c$, both gaps vanish everywhere.

In the context of $U(1)$ spin liquids in two spatial dimensions, the gap generation by CSB has the well-known interpretation as the development of Néel magnetic ordering \cite{Marston_PhysRevLett.64.1166,KIM1999130,KIM1999130,Rantner-Wen_PhysRevB.66.144501}. Indeed, we have, $\Delta_i=\sum_\alpha\langle\bar \psi_{i\alpha}\psi_{i\alpha}\rangle$. We recall that chiral condensates for a given four-component Dirac spinors are explicitly given by, $\bar \psi\psi=\psi^\dagger_R\sigma_3\psi_R-\psi^\dagger_L\sigma_3\psi_L$ (omitting all the other indices, for simplicity), where $\psi_{R}$ and $\psi_L$ are right- and left-handed spinors, respectively, and $\sigma_3$ is the third Pauli matrix. This is interpreted as the staggered magnetization, since QED3 theory is known to emerge as a low-energy effective theory of Heisenberg antiferromagnets \cite{
	Marston_PhysRevB.39.11538}. A similar physical interpretation of CSB in QED3 as an emergent staggered magnetization was considered in the past by coupling $d$-wave nodal modes in cuprates to vortices \cite{Herbut_PhysRevB.66.094504,Franz_PhysRevB.66.054535}. However, these past works use the number of fermion components as the sole parameter driving the CSB. In our case, there are two competing gauge couplings instead, although the value of $N$ still plays a role. This leads to a much richer phase structure, as shown in Fig.~\ref{Fig:PhaseDiagram}.  
          
\begin{figure}
		\includegraphics[width=7cm]{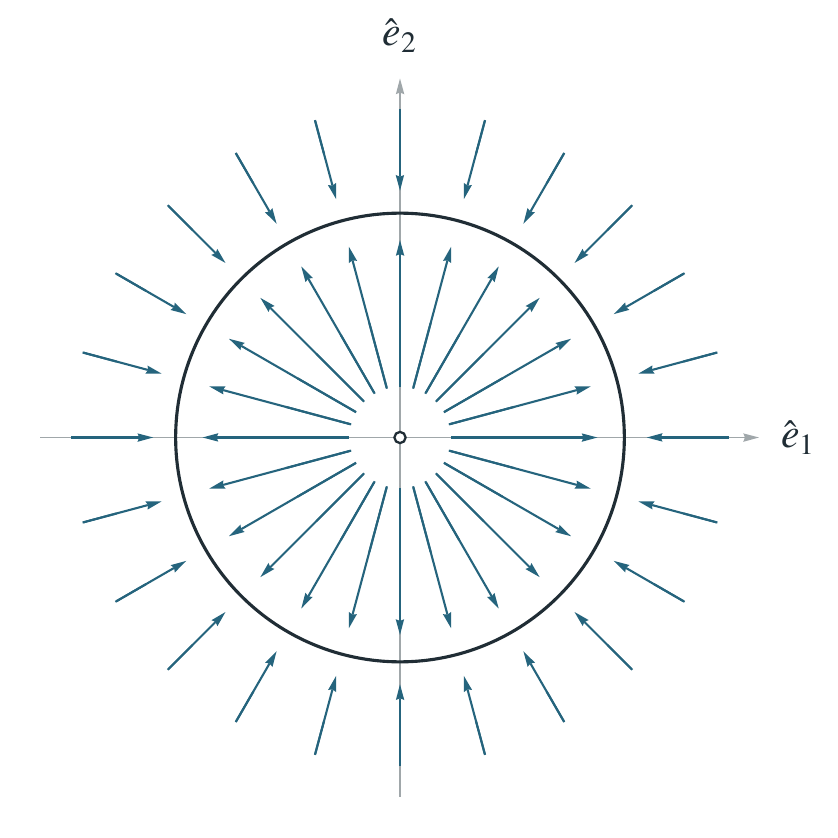}
	\caption{Schematic RG flow associated to the $\beta$ functions given by Eq. \eqref{Eq:betai}.}
	\label{Fig:RG-flow}
\end{figure}

{\it Renormalization group fixed line}--- The vacuum polarization allows us to define the dimensionless renormalized coupling constants via, 
\begin{equation}
	\label{Eq:Renorm-coupling}
	\hat{e}_i^2(p)=\frac{e_i^2|p|^{-1}}{1+\Pi(p)}.
\end{equation}
From the above we obtain the RG $\beta$ functions, 
\begin{equation}
	\label{Eq:betai}
	\beta_{\hat{e}_i^2}(\hat{e}^2_1,\hat{e}_2^2)=\frac{d\hat{e}_i^2}{dl}=\left[\frac{N(\hat{e}_1^2+\hat{e}_2^2)}{16}-1\right]\hat{e}_i^2,
\end{equation}
where $l=\ln(|p|/\Lambda)$. 
The $\beta$ functions vanish at a circular line of fixed points given by, 
\begin{equation}
	\hat{e}_{1*}^2+\hat{e}_{2*}^2=\frac{16}{N}.
\end{equation}
The RG flow is shown schematically in Fig. \ref{Fig:RG-flow}. 

We note from Fig. \ref{Fig:RG-flow} that there is no RG flow in the polar direction, the RG flow is purely radial. Furthermore, 
Eq. \eqref{Eq:Renorm-coupling} implies that $\hat{e}_2^2/\hat{e}_1^2=\kappa$ is an RG invariant. As a consequence, the $\beta$ functions can be rewritten in the form, 
\begin{equation}
	\label{Eq:beta1}
	\beta_{\hat{e}_1^2}=\left[\frac{N}{16}(1+\kappa)\hat{e}_1^2-1\right]\hat{e}_1^2,
\end{equation}
\begin{equation}
	\label{Eq:beta2}
	\beta_{\hat{e}_2^2}=\left[\frac{N}{16}\left(1+\frac{1}{\kappa}\right)\hat{e}_2^2-1\right]\hat{e}_2^2,
\end{equation}
and thus $\kappa$ can be used to parametrize the fixed circle, yielding, 
$\hat{e}_{1*}^2=16/[N(1+\kappa)]$, $\hat e_{2*}^2=\kappa\hat{e}_{1*}^2$. 
This  RG flow constitutes a genuine walking behavior \cite{Walking-JHEP} or conformality loss \cite{Kaplan_PhysRevD.80.125005}, since the flow is parametrized by an RG invariant $\kappa$ that leads to a merging of the fixed points as $\kappa\to 1$.  

We can also determine the anomalous dimensions of the Dirac fields via the expression for the wavefunction renormalizations, 
\begin{equation}
	\label{Eq:Zi}
	Z_i(p)=1+\frac{16e_i^2}{3\pi^2N(e_1^2+e_2^2)}\ln\left(\frac{|p|}{\Lambda}\right).
\end{equation}
The anomalous dimensions are given by $\eta_i=d\ln Z_i/dl$. Or, explicitly,  
\begin{equation}
	\label{Eq:etas}
	\eta_1=-\frac{16}{3\pi^2N(1+\kappa)},\quad \eta_2=\kappa\eta_1.
\end{equation} 
We thus obtain continuously varying anomalous dimensions, which evoke the BKT transition behavior. 
The anomalous dimensions of Eq. \eqref{Eq:etas} are gauge-dependent and are calculated here in the Landau gauge. 
Fixing, for instance, the Feynman gauge would lead to a positive sign of the anomalous dimensions instead of the negative one in Eq. \eqref{Eq:etas}. A gauge-independent result can be derived by inserting a Wilson line in the propagator, turning it gauge-invariant, i.e., 
$\langle\bar{\psi}_i(x)e^{ie_i\int_{x}^{y} dz_\mu a_\mu(z)} \psi_i(y)\rangle$.  As in  conventional QED3 ($\kappa=1$), this leads to a negative value of $\eta$ \cite{Gusynin_Gauge-inv-eta_PhysRevB.67.115201}. However, this procedure causes the magnitude of the anomalous dimension to differ by a factor of four. 

Considerably more significant physically is the anomalous dimension of the mass operator $O_{\bar \psi_i\psi_i}=\bar \psi_i\psi_i$, which 
represents the staggered magnetization in a spin liquid context when the chiral symmetry is broken. The value of this anomalous dimension for QED3 has been calculated using the $1/N$ expansion and is naturally gauge-independent \cite{J_A_Gracey_1992,GRACEY1993415}, since $O_{\bar \psi_i\psi_i}$ is a gauge-invariant operator. The gauge-independent result is obtained after adding  the anomalous dimension due to the mass renormalization, $\eta_{M_i}$ to the anomalous dimension $\eta_i$ of the fermions \cite{GRACEY1993415}, i.e., overall we obtain  $\eta_{\bar \psi_i\psi_i}=\eta_{M_i}+\eta_i$. 
Accounting for both coupling constants, we obtain the anomalous dimension due the mass renormalization, 
\begin{equation}
	\label{Eq:eta-Mi}
	\eta_{M_1}=-\frac{16}{N\pi^2(1+\kappa)},\quad \eta_{M_2}=\kappa\eta_{M_1}, 
\end{equation}
and hence, the mass operator anomaluous dimension,
\begin{equation}
	\label{Eq:eta-chiral}
	\eta_{\bar \psi_1\psi_1}=-\frac{64}{3\pi^2N(1+\kappa)},\quad \eta_{\bar \psi_2\psi_2}=\kappa\eta_{\bar \psi_1\psi_1}. 
\end{equation}
 A relevant physical question concerns the values of $\kappa$ corresponding to the quantum critical point. Panel (c) of Fig. \ref{Fig:PhaseDiagram} shows the gap configuration featuring an intermediate gapless phase. This occurs when $\kappa\in[\kappa_{c1},\kappa_{c2}]$, and so we have two critical end points. At $\kappa_{c1}$ one obtains $\eta_{\bar \psi_1\psi_1}(\kappa_{c1})=\eta_{\bar \psi_2\psi_2}(\kappa_{c2})=-1/3$ and  $\eta_{\bar \psi_2\psi_2}(\kappa_{c1})=\eta_{\bar \psi_1\psi_1}(\kappa_{c2})=(1/3)(1-64/\pi^2N)$. Since these critical end points are only defined for $N<N_c$, both anomalous dimensions are negative. These results have a significant impact on the behavior of the chiral susceptibility $\chi_{ij}(x)=\langle\bar \psi_i(x)\psi(x)\bar \psi_j(0)\psi_j(0)\rangle$ describing the correlation response of staggered magnetization in the gapless phase. We obtain, 
 \begin{equation}
 	\label{Eq:chis-1}
 	\chi_{11}(x)|_{\kappa=\kappa_{c1}}=\chi_{22}(x)|_{\kappa=\kappa_{c2}}\sim\frac{1}{|x|^{4/3}},
 \end{equation}
 \begin{equation}
 	\label{Eq:chis-2}
 	\chi_{11}(x)|_{\kappa=\kappa_{c2}}=\chi_{22}(x)|_{\kappa=\kappa_{c1}}\sim\frac{1}{|x|^{\frac{1}{3}\left(5-\frac{64}{\pi^2N}\right)}},
 \end{equation}
 \begin{equation}
 	\label{Eq:chi12}
 	\chi_{12}(x)\sim\frac{1}{|x|^{2-\frac{64}{3\pi^2N}}}.
 \end{equation}  
We note that the scaling behavior of susceptibility $\chi_{12}(x)$ correlating different spin species is $\kappa$-independent. 

{\it Conclusion}--- Our results identify the imbalance between gauge couplings in $U(1)$ spin-liquid mixtures as an additional control parameter of magnetic ordering. Within the leading large-$N$ and SD description, the common gauge field links two chiral instabilities whose thresholds can overlap, producing coexistence of the corresponding N\'eel orders, or leave a finite gapless interval between them. The fermion number determines which arrangement occurs, while the 
 gauge couplings' ratio $\kappa$
tunes the system through the phase boundaries. The associated fermion mass scales vanish with Miransky essential singularities. In the gapless regime, the RG-invariant coupling ratio labels a continuous family of infrared fixed points belonging to the fixed circle of the massless theory. The gauge-independent mass-operator anomalous dimensions give this structure a physical manifestation in continuously varying staggered magnetization correlations, while the exponent for correlations between the two sectors remains independent of the coupling ratio at the considered order. The same interaction imbalance therefore controls the pattern of magnetic symmetry breaking and the algebraic correlations of the gapless state, providing a framework for quantum critical behavior in systems with inequivalent spin liquid sectors.

\bibliography{CSB-QED3}

\end{document}